\documentclass[useAMS,usenatbib,usegraphicx]{mn2e}

\title[Microlensing in phase space I: Continuous limit]{Microlensing in phase space I: Continuous propagation of variability moments}

\author[Tuntsov \& Lewis]
{A.V. Tuntsov\thanks{E-mail:tyomich@physics.usyd.edu.au} 
\& G.F. Lewis \thanks{E-mail: gfl@physics.usyd.edu.au} \\ 
A29, School of Physics, University of Sydney, NSW 2006, Australia}

\begin{document}

\date{Accepted 2006 April 13. Received 2006 April 12; in original form 2005 August 31}

\pagerange{\pageref{firstpage}--\pageref{lastpage}} \pubyear{2006}

\maketitle

\label{firstpage}

\begin{abstract}

A method to calculate the statistical properties of microlensing light curves is developed. The approach follows works
by Deguchi \& Watson, Seitz \& Schneider and Neindorf, attempting to clarify the ideas involved and techniques used in the
calculations. The method is then modified to include scattering by multiple lensing planes along the line of sight and 
transition to a continuous limit of this treatment for average quantities is performed leading to a Fokker-Planck type equation. 
The equation is solved for a particular model of the random star field and microlensing effect on the flux temporal 
variability is extracted. Applications in astrophysically relevant situations are discussed.

\end{abstract}


\section{Introduction}

Gravitational microlensing of distant sources has been developed as a tool to study dark objects for more than
40 years now since pioneering works by \citet{refsdal64, refsdal66}, \citet{liebes} and \citet{byalko} (the idea
can be traced, of course, to even earlier works as listed in the Introduction of Schneider, Ehlers \& Falco (1992) monograph) 
and has attracted considerable interest since the effect was discovered in the images
of the macrolensed quasar QSO2237+0305 \citep{irwin, corrigan}. In the years leading to the discovery -- and even more after it was announced --
the researchers working in the field turned their attention to the systematic study of the statistical properties
of microlensing as it was soon realised that the number of microlenses significantly contributing to flux variations is normally large and therefore individual events in the light curves can rarely be unambiguously modelled \citep{schneiderweiss}.

In most cases of interest, gravitational microlensing can be studied in the geometric optics approximation together with
the assumption that the deflectors' gravitational field is weak and static.
The progress in this field of research has been made along a number of different complimentary directions -- both in terms
of the questions answers to which were sought and the methods used in this search. On the one hand, many important result
were obtained for the probability distribution function of the microlensing magnification value $\mu$. For example, on the 
analytic side, the large $\mu$ behaviour of this function was found \citep{peacock, schasymptotics}. However, only numerical methods --
most noticeably, the ray-shooting (Kayser, Refsdal \& Stabell 1986; Schneider \& Weiss 1987; Wambsganss, Witt and Schneider 1992) and image-tracking techniques \citep{geraint, witt} --
turned out to be capable of the determination of this function over most of its domain. These focused on extracting the
magnification PDF from many realizations of the magnification value as a function of the source position with respect to
the microlenses. Starting from a simple model of static shearless random microlens field this approach has been followed to
include the external shear and even the random proper motions of the microlenses \citep{kundic93, kundic95, schramm93}.

Perhaps less attention has been paid to a related question of correlations between the magnification values for a source placed
in different positions in the microlens field. The autocorrelation function of this kind has been studied numerically using
large representative samples of the light curves of sources tracked through magnification maps for point-like and extended
sources (Wambsganss, Paczy\'nski \& Katz 1990; Lewis \& Irwin 1996). On the analytic side, Deguchi \& Watson (1987) first calculated the variance 
of the magnification value for a Gaussian source and then Seitz and Schneider (1994) presented a general theory for the calculation
of the autocorrelation function of a static random field of stars. However, although they developed a comprehensive theory
to calculate the correlated deflection probability distribution used in calculating the autocorrelation function
its practical application turned out to be rather difficult for technical reasons. A further major step was taken by Neindorf (2003)
who found a simpler analytic expression for a restriction of correlated deflection probability Fourier transform sufficient 
to calculate the flux autocorrelation function for a circularly symmetric Gaussian source. 

One conspicuous distinction between the two classes of works mentioned above is that only the former need to solve the lens
equation. The reason here is that the quantity they are concerned with -- the magnification factor $\mu$ or the flux associated
with a given source -- is specific to a particular point in the observer's plane while the initial data to calculate it
is contained in the intensity field. Physically, the intensity 
represents the density of the photons in the phase space spanned by their coordinates
and linear momentum projections -- hence, the use of `phase space' in the title 
of our study.

Viewed in this way, solving the lens equation amounts to `deprojecting' the flux distribution
onto the intensity field; this mapping is normally one-to-many and research thus focuses on the deformations of the infinitesimal
light beams and multiplicity of solutions bringing about the caustic structure. On the contrary, the flux autocorrelation function
is defined as the average of various moments over the observer's plane. Since every light ray eventually reaches some point
in this plane one deals with the transformation of the intensity field as a whole; the conclusions are drawn from this transformed
data. There is no need to consider caustics in this case since there are no caustics in the phase space.

The purpose of the present paper is firstly, to spell out more clearly how the intensity field changes under the influence
of gravitational microlensing and how it can then be used to calculate the quantities of observational interest. We try to
identify explicitly which physical assumptions are responsible for a number of properties of this transformation which
make calculations easier and more physically meaningful. We show that the conservation of the surface brightness in
gravitational lensing -- which is an analogue of the statistical Liouville theorem -- leads to the linearity of the transformation
of the intensity n-point probability distribution functions and their moments. This leads to a slightly different description
of the gravitational microlensing which is, perhaps, most useful in analytic studies, since the dimensionality of the problem
grows rapidly with the transition from the configuration to phase space making it less amenable to numerical investigation.

We highlight how an obvious distinction between the free propagation of light rays and their deflection on the lens plane
reflects in the structure of operators performing the linear transformation mentioned above and in particular, how certain
statistical symmetries of the deflector field select a basis in which the deflection operator assumes diagonal form. Dependence
on the time coordinate is also brought in.

Another advantage of the phase space description is that virtually no effort is needed to incorporate multiple lens planes 
along the line of sight. Perhaps less indisputable but yet more appealing is the transition to a continuous limit instead
of many lens planes. We argue that in doing so one does not have to abandon the thin lens approximation when the average 
quantities such as the autocorrelation function are considered and their transformations can indeed be approximated as continuous;
this is similar to a macroscopic description of the scattering phenomena although their internal mechanics is microscopic.
After an argument in favour of the suggested approximation we obtain an equation which describes the intensity moments
propagation in space filled with randomly placed individual deflectors. 

The rest of the paper is more illustrative and rather technical in scope. We consider the propagation of the first order 
moments which are of very limited observational interest on their own but will be the basis for the study of the autocorrelation
function propagation in the companion paper. We calculate the Fourier transform of the deflection angle and potential time
delay probability distribution density and try to explicitly identify
the physical meaning of various results and techniques used. The expression obtained turns out to be weakly dependent on
the time coordinate thus reducing to a classic Katz et al. (1986) result. In addition, we find that this function is dominated
by the first nontrivial term over the entire domain of interest and illustrate how the coefficient in front of it can be obtained
in a simple alternative way.

Then the moments propagation equation with the approximate potential is solved in the case when parameters of the deflectors
are assumed constant along the line of sight. The basis of the solutions and the associated eigenvalues are found which can be
used to solve the Cauchy problem for the equation. A particular example of an isotropic source with an arbitrary spatial profile
is considered and a filter function describing different harmonic modes damping is calculated. The last section
discusses the results and their connection to observationally accessible variability timescales.

Throughout the paper we consider flat and static geometry and the results therefore can only describe local Universe $z\la 1$; 
however, this can be generalized to different universe models although the resulting differential equations will no longer
be autonomous. 

Paper II of this series \citep{paperii} presents the calculation of how the flux autocorrelation function and power spectra
are affected by gravitational lensing following the approach present in this study. The results are applied to modulation
of pulsars' emission, coherence estimates for quasi-periodic oscillations in dwarf novae and low-mass X-ray binaries and 
the interpretation of power spectra of active galactic nuclei. Microlensing-induced scatter in brightness of distant supernovae is also considered.

\section{Model}

In this section, we introduce the model used to characterize observational properties of
lensed objects. The idea is to follow the propagation of the statistical properties of the
intensity field and its moments from one lensing plane to another. Later we will
attempt a transition to the continuous limit of this treatment. The geometric optics
approximation is used and the angles involved are assumed small throughout; this is
typical for studies of most aspects of the gravitational lensing \citep{schneiderbook}.

The advantage of using the intensity -- that is, the amount of radiation energy
traversing a unit area per unit time {\it per unit solid angle} -- is that this quantity is directly
connected to the observable quantities such as flux and, by virtue of the Liouville's theorem, is conserved whenever no
absorption is involved and photon frequencies are not changed. This includes the case of gravitational microlensing where
such conservation is referred to as that of the surface brightness.

The formal connection between the intensity and the phase space density can be made
by noticing that for paraxial rays the variable component of the momentum of a photon
is proportional to the direction of the light ray, its energy is not changed between
the emission and observation points (even if this is the case, the correction
is simply the frequency ratio cubed), and the dependence of the line-of-sight 
coordinate on time is always one-to-one for such rays.

\subsection{Statistical description and the ensemble}

Now assume that we define a number of surfaces between the source and the
observer in such a way that the full knowledge of the intensity field on one of them is
sufficient to calculate it on the next one.

\begin{figure}
\hspace{0cm} \includegraphics[width=80mm, angle=0]{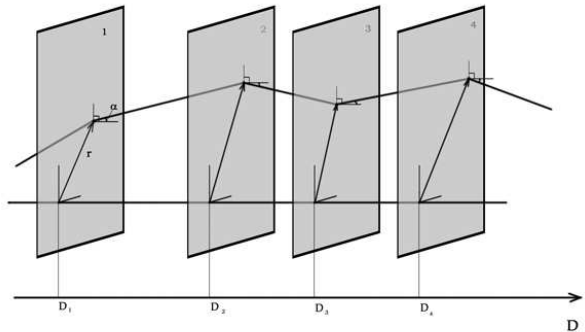}
\caption{A schematic view of a light ray, deflected by a number of parallel `lensing planes' and travelling
freely between them. The coordinates and momenta of the ray at every plane are given by ${\bf r}$ and $\balpha$, as shown
for the first plane, the positions which correspond to the same coordinate $\bf r$
in different planes are those connected by the rays perpendicular to the planes. The planes are distinguished by the depth parameter $D$, its
axis is directed towards the observer.}
\label{schemfig}
\end{figure}

The simplest choice is a number of the so
called lensing planes \citep{blandfordnarayan, nottalechauvineau, kovner87, jaroczynski} orthogonal to the line connecting the source
and the observer. The planes can be labelled with parameter $D$ starting with $D=0$ at the source (Figure~\ref{schemfig}). At this stage,
we do not include any relativity in the analysis and therefore our scope has obvious limitations. 

For the purpose given above, a sufficient way to characterize the intensity field on a
certain surface is to specify the values of the radiation intensity for every
point $\bf r$ on the surface along every direction $\balpha$ in the forward direction and,
possibly, at every instant $t$, $I({\bf r}, {\balpha}, t)$. Note that in the
light of the paragraph just before the beginning of this subsection, these coordinates 
fully parameterize the phase space of the photon dynamics $\mathcal{M}\ni ({\bf r}, {\balpha}, t)$.

Often such a complete description of the intensity field
is not available but it is possible to use a statistical description instead -- e.g.,
the $n$-point probability distribution function
$P_n(\{I\}_n, \{ r, \alpha, t\})$ defined
as the probability to observe intensity values less than $I_1$, ..., $I_n$ at the respective
coordinates $({\bf r}_1, \balpha_1, t_1), ..., ({\bf r}_n, \balpha_n, t_n)$. (We denote
these arrays of $I$ and $({\bf r}, \balpha, t)$ as $\{I\}_n$ and $\{r, \alpha, t\}_n$ to avoid
unwieldily long notation.)

The probabilities here can be defined as a fraction of appropriate realizations of the intensity field
within a certain ensemble. One can think of two major ways to define such an ensemble in the context
of microlensing.

Firstly, we can consider a number of identical at $D=0$ realizations of the intensity field in faraway patches
of the sky. As the light rays propagate towards the observer they encounter different realizations of microlenses
and the intensity field realizations are subject to individual transformations on their way. As a result,
$I({\bf r}, {\balpha}, t; D)$ is a random function with parameter $D$. Our aim is to understand
how its distribution functions $P_n$ change with $D$.

Alternatively, one can consider a single source and assume that on a sufficiently long time scale, the light rays 
emitted by the source at separate time moments encounter independent realizations of deflectors on their way 
to the observer. In this case, the elements of the ensemble can be obtained by observing the intensity field at
different times. 

An important point is that the time separation between such observations should be
long enough that the deflectors' configurations at these moments can be considered independent of each other.
In practice this means that whatever our observable is, the time separation between individual measurements
used to obtain the average of this observable must be much longer than the typical correlation time for this
quantity fluctuation due to an individual lensing plane. The second alternative can only be sensibly applied
to a statistically stationary source.


It is normally assumed that the knowledge of $n$-point distribution functions of all
orders provides the most complete statistical description of the intensity random field.
However, just a few first ones and their moments can be useful in connecting
statistical properties of the deflectors to the information that is available through
observations. 

For instance, since the angular
coordinate is normally unresolved in gravitational microlensing observations, the
quantity of interest is the flux $F$ -- the integral of the intensity over all possible
directions. Defined in this way, it represents a linear functional on the intensity field
\begin{equation}
F[I]({\bf r}, t; D)=\int\mathrm{d}^2\balpha\, I({\bf r}, \balpha, t; D)  \label{fluxdef},
\end{equation}
and provided that the values of the latter are bound, it is easy to see that the
expectation value of the flux is the same functional on the expectation value of the
intensity:
\begin{equation}
{\bf E}\, F[I]({\bf r}, t; D)=\int\mathrm{d}^2\balpha\, {\bf E}\,I({\bf r}, \balpha, t; D)=F[{\bf E}I]({\bf r}, t; D)  \label{fluxexpectation} .
\end{equation}
This is useful when we have a method to calculate the expectation value of $I({\bf r},\balpha, t)$ developed below.

It is not difficult to devise a similar linear device to obtain the expectation
value of the product of fluxes at two different points -- i.e., the flux autocorrelation
function. One would need, however, to define it on the intensity autocorrelation field
to preserve linearity and thus the simple way to calculate the expectation value.

\subsection{Linearity of the intensity field transformation}

Once an $n$-point distribution function $P_n(\{I\}_n, \{r, \alpha, t\}_n)$ has been specified on a
certain surface, its subsequent propagation is described by a linear relation. 

Indeed, assuming there is no absorption, the intensity is conserved along the ray, and the process we are dealing with is just a sort of
mixing -- the probability $P'_n$ to observe a certain configuration $\{I\}_n$ of the intensity field
at points $\{r, \alpha, t\}'_n$ on the next surface is just that $P_n$ on the initial surface at points
$\{r, \alpha, t\}_n$ connected to $\{r, \alpha, t\}'_n$ by the light rays:\
\begin{equation}
P'_n\left(\{I\}_n, \{r,\alpha, t\}_n'\right)=P_n\left(\{I\}_n, \{r,\alpha, t\}_n\right) \label{Pchange}.
\end{equation}

The case of interest is when the
mapping $\{r, \alpha, t\}_n\rightarrow\{r, \alpha, t\}'_n$ is itself of random nature measured by the
probability density $p_n\left(\{r, \alpha, t\}_n\rightarrow\{r, \alpha, t\}'_n\right)$ for light rays
starting at $\{r, \alpha, t\}_n$ on the first surface to get to $\{r, \alpha, t\}'_n$ on the next one.
Then, the transformation law for $P_n$ is given simply by the total probability formula:
\begin{eqnarray}
\lefteqn{P'_n\left(\{I\}_n, \{r,\alpha, t\}_n'\right)=}\label{Ptransform}\\
\lefteqn{\int\mathrm{d}\{r,\alpha, t\}_n\, p_n\left(\{r,\alpha, t\}_n\rightarrow\{r,\alpha, t\}_n'\right) P_n\left(\{I\}_n, \{r,\alpha, t\}_n\right)} , \nonumber
\end{eqnarray}
where $\mathrm{d}\{r, \alpha, t\}_n\equiv\prod\limits_{i=1}^n\rmn{d}^2{\bf r}_i\rmn{d}^2\balpha_i\rmn{d}t_i$ was used for brevity.

Strictly, this expression is only valid when no point $\{r,\alpha, t\}_n'$ is shadowed -- that is, all possible mappings in the backward direction lead
to a certain point $\{r,\alpha, t\}_n$ at the preceding plane; this is normally warranted by the symmetry of $p_n$ and its normalization. Even if this
condition is not satisfied, the results are still valid for monomial moments of the intensity.

%
%

The `transition probabilities' $p_n$ are independent of $I$ and therefore the same transformation
law is valid for the probability densities $\mathrm{d}^nP/\mathrm{d}\{I\}_n$ 
and its various intensity moments -- e.g.,
the intensity $n$-point correlation function:
\begin{equation}
\lefteqn{M_n\left(\{r, \alpha, t\}_n\right)\equiv\int I_1\cdot ...\cdot I_n\,\mathrm{d}^nP_n\left(\{I\}_n, \{r, \alpha, t\}_n \right)}. \label{Mdef}
\end{equation}

Therefore, the set of all such intensity moment functions on $\mathcal{M}^n$ naturally
acquires the strutcure of a linear space  $\mathcal{V}(\mathcal{M}^n)$; this is useful
because the properties of the transformations applied to the intensity moments
often suggest particularly convenient bases in which to represent $M_n$.

\subsection{Deflection and focusing}

From the intensity field point of view, light propagation in the presence of gravitational lensing involves two rather distinct kinds of transformations.

The curved part of the light ray trajectory is normally very small compared to other important length scales
of the problem and one adopts the so-called thin lens approximation \citep{pynebirkinshaw} -- the light rays are assumed to
be simply broken on the plane by a certain angle depending on the deflector configuration, they also receive certain time delays; we will call this deflection and denote the appropriate
linear operator $\hat\Phi$. Between such breakages, the light rays propagate freely, converging or diverging depending on the angle between them;
this will be termed focusing, $\hat{F}$.

In the thin lens approximation (sometimes termed just the gravitational lensing approximation), the spatial coordinates of the light rays 
do not change when a deflection occurs, only the angles are modified and time delays are experienced. In focusing, one has an opposite situation. This reflects in the structure 
of the kernels ($p_n$ in notation of~(\ref{Ptransform})) of linear operators which describe the two transformations. Deflection should be proportional to identity in its spatial part,
i.e. $\Phi\sim \delta\left(\{r\}_n, \{r\}_n'\right)$. The same applies to the angular part of the focusing: $F\sim\delta\left(\{\alpha\}_n, \{\alpha\}_n'\right)$.

Another major difference between deflection and focusing is that only the former can be of random nature. The result
of focusing over some distance $D$ is always the change of ${\bf r}$ to ${\bf r}+D\balpha$ and of ${t}$ to
$t+D(1+\alpha^2/2)/c$ and since the values of the intensity themselves do not change, the focusing operator $\hat{F}(D)$
reads
\begin{eqnarray}
\lefteqn{\hat{F}\left(D\right)M\left(\{r,\alpha, t\}_n\right)=} \label{focusing} \\
& &  M\left(\{r-\alpha D, \alpha, t-(1+\alpha^2/2)D/c\}_n\right) \nonumber,
\end{eqnarray}
or, for sufficiently smooth functions $M(\{r, \alpha, t\})$,
\begin{equation}
\hat{F}\left(D\right)=\exp\left[-D\sum\left(\balpha_i\cdot\partial_{{\bf r}_i}+c^{-1}\left(1+\alpha_i^2/2\right)\partial_{t_i}\right)\right]\label{focusingdiff}
\end{equation}
with $i$ running through all rays from $1$ to $n$.

On the contrary, deflection may contain both the deterministic and random parts;
because of the linearity, they can be applied separately. A simple but important
property of the deflection comes from the assumption that the angles $\alpha$ are small. Taken with the thin lens approximation,
this means that all rays are deflected by the same amount regardless of incidence angle $\alpha$. Therefore
the probability density $p_n$ for $\hat\Phi$ can depend on the difference between $\alpha$ and $\alpha'$ only.
This symmetry reflects in the fact that $\hat\Phi$ assumes a diagonal form in an eigenbasis of this transformation group
-- e.g., Fourier harmonics $\exp[-\mathrm{i}\btau\cdot\balpha]$, where it is especially simple to see what the 
coefficient in front of the $\delta$-function is.

For the deterministic part -- deflection $\bbeta_d({\bf r}, t)$ and time delay $s_d({\bf r}, t)$ due to a known deflector -- $\balpha'=\balpha+\bbeta_d$, $t'=t+s_d$ and therefore
\begin{eqnarray}
\lefteqn{\Phi_d\left(\{r, \tau, t\}_n, \{r, \tau, t\}'_n \right)=\delta(\{t\}_n+\{s\}_n, \{t\}_n') } \label{phimatrix} \\
& &\hspace{-2mm} \times\exp\left[\mathrm{i}\sum\btau_i\cdot\bbeta_d({\bf r_i}, t_i)\right]\delta\left(\{\tau\}_n,\{\tau\}'_n \right) \delta(\{r\}_n, \{r\}_n').\nonumber
\end{eqnarray}
In the case of an unknown deflector, one finds the average result of deflection using the expectation value of the exponential factor (${\bf E}\left(\hat\Phi M\right)=\left({\bf E}\hat\Phi\right) M$):
\begin{equation}
\Phi(\{r\}_n, \{\tau\}_n, \{t\}_n, \{t\}_n')\sim {\bf E}\exp\left[\mathrm{i}\sum\btau_i\cdot\bbeta({\bf r}_i, t_i)\right] \label{Phiexpectation}
\end{equation}
-- i. e., $\bbeta$ characteristic function, Fourier transform of its distribution function density $p(\{\bbeta\}; \{ r\}, \{t\})$. 

Note that this trick does not work for time delays: since the addend $s$ to $t$ depends on $t$ itself, $\exp[\mathrm{i}\omega\,s]$-like
factor would not come out of the integral as a factor when making the Fourier transform in $t$. However, when the expectation value ${\bf E}\exp[\mathrm{i}\omega\,s]$
is independent of $t_i$ for all $i=\overline{1, n}$, $\hat\Phi$ assumes a similar form in its temporal part. This is the case for a statistically 
stationary deflector when only one-ray statistics are considered. This problem, $n=1$, is considered below.


Of our prime interest are deflections due to {\it a priori}
unknown distribution of (micro)lenses and we will ignore the deterministic part for the rest of this study. 

\subsection{Multiple lens planes and the continuous limit}

The above approach, in slightly different expression and with a different emphasis, has been used to study the deflection
statistics by a single deflecting plane (e.g., \citealt{katz, neindorf}). Little change is needed to apply it to
a number of different planes when the light rays are significantly deflected a number of times on their way from
the source to the observer -- one applies the deflection and focusing transformations a number of times
with appropriate parameters thus implementing the multiple lens plane approximation within this approach:
\begin{equation}
M_O=\hat{F}(D_{Ok})\hat\Phi_k\hat{F}(D_{k, k-1})...\hat{F}(D_{2 1})\hat\Phi_1\hat{F}(D_{1S})M_S, \label{multipleplanes}
\end{equation}
where $M_O$ and $M_S$ are the average intensity at the observer and the source and $D_{j, j-1}$ is the distance between the $j$-th and $(j-1)$-th lensing planes.

Focusing operator is diagonal in spatial wavevector ${\bf k}$, incidence angle $\balpha$ and temporal frequency $\omega$ representation, as can be seen
from~(\ref{focusingdiff}). If one disregards potential time delays, the deflection operator assumes diagonal form in ${\bf r}$, $\btau$, $\omega$ basis.
The conversion between the two representations can be efficiently performed numerically which suggests using the formulation presented for numerical
studies of multiple lens planes case.

Our focus, however, is different. When one deals with a 3-D distribution of lenses such as a cosmological population
of primordial black holes, CDM haloes or dense gas clouds, this somewhat arbitrary `slicing' of 3-D space into a number
of lensing planes seems a little unnatural. In addition, since only limited information about the distribution of such
deflectors is normally available (or assumed), the main interest is general behaviour of various observables and their
averages as the light propagates through the Universe. We are therefore naturally lead to a three-dimensional problem.

The complexity of the full-scale problem is enormous and only few attempts to attack it directly have been made.
There are two main reasons for this. First is the need to abandon the thin lens approximation -- one cannot describe
a gradual change in light ray momenta within this model \citep{pynebirkinshaw, continuouslimit}. The second is also related to the thin lens approximation --
when the deflection probabilities are computed, the correlations between the deflectors belonging to different planes
would have to be considered.

However, when it comes to the statistical average of the quantities of interest, the thin lens
approximation can still be valid for two reasons. Firstly, since the incidence angles are very small, 
deflections due to different individual lenses simply add up. Secondly, in astrophysical applications, a typical scale over which moments~(\ref{Mdef}) change appreciably is much 
larger than the light ray curvature radii. Therefore the statistical averages can be taken on `statistically' 
infinitesimal  $\mathrm{d}_{\mathrm{phys}}D$ scales which are much larger than the scale relevant to elementary deflections.
 
Since the deflection angles are additive, averaging of $\mathrm{d}_{\mathrm{phys}}\hat\Phi$ reduces to the use of the thin lens approximation,
while the commutator $\left[\mathrm{d}_{\mathrm{phys}}\hat\Phi, \hat{F}(\mathrm{d}_{\mathrm{phys}}D)\right]$, which quantifies the error
introduced, is of the second order in $\mathrm{d}_{\mathrm{phys}}D$.
From now on, when some changes in the intensity distribution or moments are mentioned, the averages over the same physically 
small distances will be actually meant; the subscript `phys' in differential will also be dropped.

The above observation can also be used when computing the deflection probabilities -- whenever the scale of correlations
between individual lenses is small compared to $\mathrm{d}D$, their distribution can be considered effectively
two-dimensional. The model distribution used in the next section assumes that no correlation at all is present
and all individual deflectors are distributed independently. We are not aware of any similar calculations which would
take the correlations into account.

We have thus put some additional meaning in the thin lens approximation and can now obtain a differential equation
to describe the continuous propagation of the intensity field moments. Indeed, the changes in the intensity field {\it averaged over
the ensemble} are small on a small run $\mathrm{d}D$. Therefore we can expand $\hat{F}$ and $\hat\Phi$ operators
in series of $\mathrm{d}D$. The zeroth order term in this series is identity, and therefore, to the first order in $\mathrm{d}D$,
\begin{equation}
\frac{\partial M}{\partial D}=(\frac{\mathrm{d}\hat{F}}{\mathrm{d}D} + \frac{\mathrm{d}\hat\Phi}{\mathrm{d}D})M
\label{abst}
\end{equation}
-- a result of a typical Fokker-Planck structure $\partial_t\rho=\hat{L}\rho$.

%
For independently distributed deflectors, the only extensive parameter describing
the distribution is the total number of the deflectors $N$. Since the two otherwise similar transformations
with parameters $N_1$ and $N_2$ applied successively should produce the same result as a single transformation with the pooled parameter: 
$\hat\Phi(N_1)\hat\Phi(N_2)\equiv\hat\Phi(N_1+N_2)$, 
there exists an operator $\hat\varphi$ such that:
\begin{equation}
\hat\Phi\left(N\right)=\exp[N\hat\varphi] \label{varphidef}
\end{equation}
and since the number of deflectors in a layer of depth $\mathrm{d}D$ is proportional to $\mathrm{d}D$, one has
\begin{equation}
\frac{\mathrm{d}\hat\Phi}{\mathrm{dD}}=\left.\nu||\mathcal{L}||\frac{\mathrm{d}\exp[N\hat\varphi]}{\mathrm{d}N}\right|_{N=0}=\nu||\mathcal{L}|| \hat\varphi \label{dPhidD}
\end{equation}
where $\nu$ is the space number density of the deflectors, $||\mathcal{L}||$ is the area of the lensing plane $\mathcal{L}$ and the operator $\hat\varphi$ only
depends on the individual properties of deflectors.

In the case of the first order moments (or the 1-point intensity distribution function) the equation simplifies
significantly when one assumes (statistical) stationarity of the deflectors. In this case, the expectation
${\bf E}\exp[\mathrm{i}(\btau\cdot\bbeta+\omega\,s)]$ is independent of $t$ and $\hat\varphi$ assumes diagonal form in ${\bf r}$, $\btau$, $\omega$ basis.

There is a certain trade-off between the simplest forms of the focusing and
deflection -- one could instead leave the dependence on angles to make the focusing
algebraic but that would leave the deflection an integral operator which is harder to deal with. In addition,
many properties of the solution will be dictated by the form of the focusing which does not depend on the deflector
properties and distribution; we therefore opt for ${\bf r},\btau,\omega$ coordinates. 

Changing to this basis 
and evaluating $\mathrm{d}\hat{F}/\mathrm{d}D$ at $D=0$ using~(\ref{focusingdiff}) one arrives at a Schr\"odinger-like 
equation\footnote{We drop $\omega/c$ term which describes the fiducial light-travel time and is eliminated 
by the transformation $M\rightarrow M\exp[\mathrm{i}\omega D/c]$.}:
%
\begin{eqnarray}
\lefteqn{\mathrm{i}\partial_DM=\frac{\omega}{2c}\,\partial^2_{\btau}M-\partial_{{\bf r}}\cdot\partial_{\btau} M+\mathrm{i}VM}  \label{sme} 
\end{eqnarray}
with a complex potential $\mathrm{i}V({\bf r},\btau, \omega)\equiv\mathrm{i}\mathrm{d}\Phi({\bf r},\btau, \omega)/\mathrm{d}D$:
\begin{eqnarray}
\lefteqn{V({\bf r},\btau, \omega)=\nu||\mathcal{L}||\frac{\mathrm{d}}{\mathrm{d}N}\ln\bigl\{{\bf E}_N\exp\left[\mathrm{i}\left(\btau\cdot\bbeta({\bf r})+\omega\, s({\bf r})\right)\right]\bigr\} }  \label{potentialdef} 
\end{eqnarray}
where the subscript $N$ at the expectation symbol ${\bf E}$ means that functions $\bbeta$ and $s$ are computed for the distribution with the (average) number of deflectors $N$; the derivative is evaluated at $N=0$.

\section{Deflection probabilities for one ray}

We employ the Markov-Holtzmark-Chandrasekhar method (e.g., \citealt{chandrasekhar43}) to calculate the deflection potential $V$ due to randomly and
independently distributed deflectors. The essence of this approach is in the fact that the deflection operator for one ray is diagonal 
in ${\bf r}$, $\btau$, $\omega$ basis for a statistically stationary deflecting plane, as explained above.

Therefore, transformation due to exactly $N$ deflectors is just the $N$-th power of transformation due to one deflector:
\begin{equation}
p\left(N; {\bf r}, \btau, \omega\right) = \left(p\left(1; {\bf r}, \btau, \omega \right)\right)^N . \label{MHCh}
\end{equation}
This result has been used by a number of authors in gravitational lensing studies (Katz~et~al. 1986; \citealt{deguchi, neindorf}).

Since there is no control over the number of deflectors $N$ in the layer, and especially since we consider a `physically' infinitesimally 
thin layer $\mathrm{d}D$, the number $N$ is allowed to fluctuate. Assuming Poissonian distribution for this quantity with average $\bar{N}$,
the average transformation reads \citep{neindorf}:
\begin{eqnarray}
\lefteqn{p_{\bar{N}}\left({\bf r}, \btau, \omega\right)=\exp[\bar{N}\left(p\left(1; {\bf r}, \btau, \omega\right)-1\right)] \, . } \label{poissonp} 
\end{eqnarray}
and using~(\ref{potentialdef}) one has for the deflection potential:
\begin{eqnarray}
\lefteqn{V({\bf r}, \btau, \omega)=\nu||\mathcal{L}||\left(p\left(1; {\bf r}, \btau, \omega\right)-1\right)} \label{Vfromp1}\\
& & \hspace{8mm} =\nu||\mathcal{L}||\bigl\{ {\bf E}\exp\left[\mathrm{i}\left(\btau\cdot\bbeta_i+\omega\,s_i\right)\right] - 1\bigr\},  \nonumber
\end{eqnarray}
where $\bbeta_i$ and $s_i$ are the deflection angle and the potential time delay due to a single deflector which depend on its individual properties only;
the expectation is taken as the average with respect to all these properties.

An individual deflector can be characterized by the above two functions $\bbeta_i\left({\bf r} - {\bf r}'\right)$
and $s_i\left({\bf r} - {\bf r}'\right)$ that, in the deterministic case, describe the deflection angle and potential
time delay observed at ${\bf r}$ due to a deflector placed at ${\bf r'}$\footnote{In gravitational lensing $\bbeta_i$ is, up to a factor, 
a gradient of $s_i$, but this will not be used here.}. They depend on other deflector parameters as well,
for instance, they must be proportional to its mass $m$. In the simplest case of a point-like mass
$\bbeta({\brho})=-m\brho/\rho^2$ and $s(\brho)=-m/c\ln\rho/\rho_0$ for $\rho$ greater than a few gravitational radii
$m/2$, $\rho_0$ is a constant; it is convenient to choose it in such a way that the average potential time delay is
zero (Schneider~et~al. 1992).

In this study we restrict ourselves to the simplest case of a point-like mass; the only parameter $m$ is fixed so that the expectation value 
is calculated as the average with respect to the deflector position in the lensing plane $\mathcal{L}$:
\begin{equation}
{\bf E}\,f=\int\limits_{{\bf r}'\in\mathcal{L}}\mathrm{d}^2{\bf r}'\,\phi({\bf r}')f({\bf r}') \,,\label{expectationphi}
\end{equation}
where $\phi$ is the distribution function density of ${\bf r}'$.

It is most natural to assume that individual deflectors are uniformly distributed in $\mathcal{L}$: $\phi(\brho)=||\mathcal{L}||^{-1}$,
so that the resulting potential is independent of ${\bf r}$ in the limit of infinite lensing plane $\mathcal{L}$. 

However, the radial integration in~(\ref{expectationphi}) for $r^{-1}$ exponent diverges both at infinity and at zero. The latter limit poses
no problem because the integration should be truncated at some finite distance from the deflector $\epsilon=\mathcal{O}(m)$
-- the light rays cannot pass through if shot too close to the lens. As for the outer integration limit, no satisfactory
explanation appears in the literature for putting a certain cut-off. The discussion normally revolves around the fact
that all real distributions have some physical scale associated with them -- e.g., the size of the galaxy the deflectors
belong to. 

However, for a finite lens plane, the boundary effects start to matter, and what is difficult to secure is
$p$'s independence of ${\bf r}$. Standard derivations (Katz~et~al. 1986; Schneider~et~al. 1992) rely heavily on observation point ${\bf r}$
being the symmetry centre of the distribution and it is easy to check that
once this symmetry is broken $p$ starts to depend on ${\bf r}$; moreover, this dependence does not vanish
in the limit of an infinite $\mathcal{L}$.

This problem appears to be similar to standard cosmological paradoxes involving the infinite distribution of mass and
light in the Universe and the absence of the satisfactory solution to it might also be an indication that the consistent
solution can only be found within fully relativistic approach. However, as the latter is not available at the moment,
we will use a standard way out: to imagine that $\mathcal{L}$ is symmetric enough to make ${\bf r}$ dependence vanish
locally, calculate $p$ for such a distribution and then choose the lens spatial dimension in a way appropriate for
the problem. This is expected to capture the general behaviour of the intensity moments. However, one should keep in
mind limitations inherent in this derivation.

Armed with this knowledge, we calculate $p(\btau, \omega)=p(1; {\bf r}, \btau, \omega)$ for a deflector distributed in a disc
of radius $R$ for observation point ${\bf r}={\bf 0}$. That is, we choose the distribution
\begin{equation}
\phi(\brho)=\frac{1}{\pi R^2}, \hspace{1cm} \epsilon\le\rho\le R \label{phidisc}
\end{equation}
and zero otherwise; $\epsilon\ll R$ is assumed, so that $\phi(\brho)$ is still normalized.
We use $\bbeta_i(\brho)=-m\brho/\rho^2$ and $s_i(\brho)=-m/c\ln\rho/\rho_0$, $\rho_0$ parameter will be chosen shortly.

To calculate $p(\btau, \omega)$ we expand the exponent in its Taylor series and then integrate it term by term:
\begin{equation}
p(\btau, \omega) = {\bf E}\bigl\{1+\mathrm{i}\left(\btau\cdot\bbeta_i+\omega\,s_i\right) - \frac{1}{2}\left(\btau\cdot\bbeta_i+\omega\,s_i\right)^2+...\bigr\}. \label{pseries}
\end{equation}
The resulting series converges for $\phi$ defined by~(\ref{phidisc}), and therefore equals the initial integral.

The actual integration is done in Appendix~\ref{potentialappendix} and the result is presented below. However, the expansion above is quite useful 
in clarifying the physical meaning of the potential function $V$, proportional to $p-1$. The first term is the normalization, integral of $p$ over all 
$\bbeta$ and therefore $V$ expansion starts from zero. 

The second term contains average $\btau\cdot\bbeta$ and $\omega\,s$. The former has to be zero
because of the symmetry: all the directions of $\bbeta$ are equally likely; for the same reason all other terms containing odd powers of $\btau$ vanish.
As for the average potential time delay ${\bf E}s$, we can use our freedom in choosing $\rho_0$ in $s_i(\brho)$ to make the average potential time delay zero
as there is no way we can measure this delay directly.

Therefore, the expansion of $V$ starts with terms proportional to the squares of $\btau$ and $\omega$, the cross-term in the square of the sum vanishes 
because it contains an odd power of $\bbeta$. In fact, the quadratic term in $\btau$, proportional to the dispersion of $\bbeta$ dominates $V$, the correction
is a slowly varying function of order $\ln^{-1}\Lambda$, where $\Lambda\equiv R/m$, and one can write, for $\omega=0$:
\begin{eqnarray}\
\lefteqn{p(\btau)\approx 1 - \frac{\tau^2}{4}\,{\bf E}\bbeta^2}, \label{secondterm}
\end{eqnarray}
where an extra $1/2$ is the average value of $\cos^2(\bbeta, \tau)$. An analogous interpretation for $\omega\not = 0$ can hardly be provided because the deflection angle and potential time delay 
are strongly correlated -- in fact, $\bbeta$ uniquely determines $s$.

The average square of $\bbeta$ (which is also its dispersion) can be calculated in a simple way by inverting $\bbeta(\brho)$
to obtain the distribution density of the deflection angle
\begin{equation}
p(\bbeta)=\phi(\brho(\bbeta))\left|\frac{D(\brho)}{D(\bbeta)}\right|=||\mathcal{L}||^{-1}\frac{m^2}{\bbeta^4}. \label{bbetadistribution}
\end{equation}
Observing that $\beta$ ranges from $\beta_{\mathrm{min}}\sim m/R$ to $\beta_{\mathrm{max}}\sim 1$ one uses the above distribution to
 calculate
\begin{equation}
{\bf E}\bbeta^2=2\pi\frac{m^2}{||\mathcal{L}||}\ln\Lambda \label{bbetasquaredmean}
\end{equation}
and at $\omega\rightarrow 0$ one has
\begin{equation}
V(\tau)=-\frac{\pi}{2}\nu m^2\tau^2\ln\Lambda .\label{Vomegazero}
\end{equation}

More accurate calculation, performed in Appendix~\ref{potentialappendix}, results in the following deflection potential:
\begin{eqnarray}
\lefteqn{V(\btau, \omega)=\pi\nu\left[R^2\Delta f_\omega\left(-\frac{m\omega}c, \Lambda\right) + m^2 f_\omega\left(-\frac{m\omega}c, \Lambda\right)\right] }\label{Vdisc} \\
& & \hspace{-7mm} -\frac{\pi}2\nu m^2\tau^2\ln\Lambda\left[\frac{\exp(\mathrm{i}\omega/\omega_0)-1}{\mathrm{i}\omega/\omega_0}+\frac{\exp(\mathrm{i}\omega/\omega_0)}{\ln\Lambda}f_\tau\left(-\frac{\tau^2}{4}\right)\right]. \nonumber
\end{eqnarray}
Functions $f_\tau$, $f_\omega$ and $\Delta f_\omega$ are defined in Appendix~\ref{potentialappendix} and $\omega_0\equiv c/(m\ln\Lambda)$.

The first line in the above expression, which is independent of $\btau$, describes some additional spread in light ray arrival times 
due to the variance of potential time delays within the ensemble. It is easily taken into account and will be discussed in greater detail when solving 
the equation~(\ref{sme}) in the next section.

For the second line, which represents the $\btau$-dependent part $V'$ of the deflection potential $V$, one can notice that for a solar-mass lens and 
$\ln\Lambda\sim 10 - 10^2$, the characteristic frequency $\omega_0\sim 5\times(10^2 - 10^3)\,\mathrm{s}^{-1}$. This is much
higher than typical observationally measurable frequencies in astronomy, and in the leading order, one obtains
\begin{equation}
V'(\btau)\approx-\frac\pi2\nu m^2\tau^2\ln\Lambda\left[1+\frac{1}{\ln\Lambda}f_\tau\left(-\frac{\tau^2}{4}\right)\right]; \label{Vprimeleadingorder}
\end{equation}
This approximation for $V'$ is independent of the frequency $\omega$. Physically, neglecting this dependence 
means ignoring the potential time delays altogether, as can be seen from~(\ref{potentialdef}). 

This can indeed be done due to the logarithmic dependence of the potential time delay $s$ on the
impact parameter $\rho$: $\rho\sim\exp[-s/2t_m]$, $t_m=m/2c$ is the gravitational radius light crossing time. 
The average time delay is unobservable and we remove it by choosing appropriate $\rho_0$; at
the same time, close encounters are rare for realistic deflector densities, and one can safely neglect all deflectors
except the dominant, closest one. The distance to this deflector $\rho$ in the layer of depth $d$ is distributed according 
to the Poisson's law: $P(\rho)=1-\exp[-\pi d\nu \rho^2]$ and therefore potential time delay distribution density is 
\begin{equation}
\frac{\mathrm{d}P(\rho(s))}{\mathrm{d}s}\approx\frac{\pi\nu d\rho^2_0}{t_m}\exp\left[-\pi\nu d\rho^2_0e^{-s/t_m}-s/t_m\right], \label{delaydensity}
\end{equation}
-- i.e., it is exponentially suppressed for $s$ greater than a few $t_m\approx 10^{-5}(M/M_\odot)\,\mathrm{s}$.

In addition, because function 
$f_\tau$ is very weakly dependent on its argument (see Appendix~\ref{potentialappendix}) and supressed by $\ln^{-1}\Lambda$ factor in~(\ref{Vdisc}), this dependence will also
be neglected for the time being. As a result, $V'$ reduces to~(\ref{Vomegazero}) and the equation~(\ref{sme}) is formally similar to the Schr\"odinger
equation for harmonic oscillator with a complex constant.

\section{Solution of the equation}
We are now ready to formulate the problem properly. We seek a solution of the equation~(\ref{sme}) with a potential given by~(\ref{Vdisc}) and
initial conditions
\begin{equation}
M({\bf r}, \btau, \omega)\left|_{D=0}\right.=M_0({\bf r}, \btau, \omega) \label{initialconditions}
\end{equation}
When the deflectors' distribution is uniform, the potential is independent of ${\bf r}$ and things simplify further by changing into
Fourier domain in ${\bf r}$ space:
\begin{eqnarray}
\lefteqn{\partial_D M({\bf k}, \btau, \omega)={\bf k}\cdot\partial_\btau M({\bf k}, \btau, \omega) -\frac{\mathrm{i}\omega}{2c}\partial^2_\btau M({\bf k}, \btau, \omega)} \label{smek} \\
& & \hspace{15mm} +V(\btau, \omega)M({\bf k}, \btau, \omega) \nonumber
\end{eqnarray}
%
Therefore, equations for different ${\bf k}$,$\omega$-modes $\xi_{{\bf k}\omega}(\btau)$ decouple for such potential, leaving ${\bf k}$ and $\omega$ as
parameters of the equation. This starts the procedure of variable separation performed in Appendix~\ref{solutionappendix}.

Before this, however, let us observe one further simplification of the problems which arises from
a particular form of the potential. Indeed, the term proportional to the first line of~(\ref{Vdisc}), that is independent of $\btau$, 
can be eliminated from the equation by making a change of the dependent variable:
\begin{equation}
\xi_{{\bf k}\omega}(\btau)=\xi'_{{\bf k}\omega}(\btau) w(\omega, D) \label{xipgeneral} \,
\end{equation}
with
\begin{equation}
w\left(\omega, D\right)\equiv\exp\left[\pi\int\limits_0^D\mathrm{d}D' \nu\left(R^2\Delta f_\omega+m^2 f_\omega\right)\right] \label{wdef} ,
\end{equation}
which leaves only the $\btau$-dependent part of the potential $V'$
in the equation for $\xi'$.

The physical meaning of the transformation~(\ref{xipgeneral}) can be understood using the Fourier convolution theorem.
Indeed, multiplication of Fourier transforms corresponds to convolving the $t$-represented moment $M$ with a `spread function' $w(t-t')$
whose Fourier transform is given by~(\ref{wdef}). 

Therefore, when one considers a `synchronous' source, in which temporal variations 
only modulate a given intensity distribution:
\begin{equation}
M_0({\bf k}, \btau, \omega)=A(\omega)I({\bf k}, \btau) \label{modulation}
\end{equation}
modification~(\ref{xipgeneral}) applies to the amplitude factor $A$ only. As a result, evolution of $M$ is split into
a non-trivial part caused by deflections and described by equation on $\xi'$ with potential $V'$, and the overall spread 
in light-ray arrival times, which comes from the variance in potential time delays within the deflector ensemble. The latter part
is described by~(\ref{wdef}), which is discussed in greater detail in Appendix~\ref{spreadappendix}.


The resulting problem for $\xi'_{{\bf k} \omega}$
\begin{equation}
\partial_D\xi'={\bf k}\cdot\partial_\btau\xi'-\frac{\mathrm{i}\omega}{2c}\partial^2_\btau\xi'+V'\xi' \label{xipproblem}
\end{equation}
is solved in Appendix~\ref{solutionappendix} using separation of variables for the $D$-independent distribution of microlenses under assumption
that $f_\tau$ term can be neglected in $V'$, as explained in the previous section.

A complete eigensystem of solutions $M$ and corresponding $D$-eigenvalues $\lambda$ of~(\ref{smek}) is built which provides the solution to the problem
presented in the beginning of this section for any initial conditions~(\ref{initialconditions}) as a sum (integrals with respect to ${\bf k}$, $\omega$ 
in a non-compact case):
\begin{eqnarray}
\lefteqn{M({\bf r}, \btau, t; D)=} \label{finalsolution} \\
& & \sum\limits_{{\bf k} n l\omega} w(\omega, D) \left(M^\dag_{{\bf k} n l\omega}, M_0\right)\mathrm{e}^{\lambda_{{\bf k} n l\omega}D} M_{{\bf k} n l\omega}({\bf r}, \btau, t)  \nonumber
\end{eqnarray}
with the basis functions
\begin{eqnarray}
\lefteqn{M_{{\bf k} n l\omega}({\bf r}, \btau, t)=} \label{firstordereigenfunctions}\\
& & (2\pi)^{-2}\exp\left[-\mathrm{i}\left({\bf k}\cdot{\bf r}+\omega t+n\chi+\frac{c}{\omega}{\bf k}\cdot\btau \right)\right]\zeta_{\omega n l}(\tau) , \nonumber
\end{eqnarray}
and associated eigenvalues
\begin{equation}
\lambda_{{\bf k}\omega n l}=-\frac{\mathrm{i}c}{2\omega}{\bf k}^2+\mathrm{i}\frac{\omega}{c}\sqrt{\mathrm{i}\kappa(\omega)}\left(1+|n|+2l\right), \label{firstordereigenvalues}
\end{equation}
where the `oscillator constant' $\kappa$ is defined in analogy with QM harmonic oscillator:
\begin{equation}
\kappa(\omega)=\pi\nu\frac{m^2c}\omega\ln\Lambda . \label{kappappbody}
\end{equation}
so that at low frequencies $\omega\ll\omega_0$ one has $2c V'/\omega\approx\kappa\tau^2$.

Function $\zeta_{n l \omega}(\tau)$ is the eigenfunction corresponding to $l$-th radial mode of~(\ref{xipproblem}), it is defined in Appendix~\ref{solutionappendix}; 
index $n$ denotes angular modes. It is convenient here to present the solution as a function of $\btau$ (rather than $\balpha$) because the flux corresponds to value of intensity moment at $\btau=0$.

The scalar product $\left( \cdot , \cdot \right)$ used above to project the initial conditions onto the eigenfunctions is defined in the standard way (first factor is complex-conjugated in the integral) in $\mathcal{V}(\mathcal{M})$.
Because the potential $V'$ is complex-valued, Hermitean conjugates $\zeta^\dag_{n l \omega}$ and $\zeta_{n l \omega}$ are complex conjugates of each other; in this
way the orthogonality relation~(\ref{orthogonality}) holds, as explained in the Appendix~\ref{solutionappendix}. The conjugates $M^\dag_{{\bf k} n l \omega}$ to $M_{{\bf k} n l \omega}$ 
differ only in the use of $\zeta_{n l \omega}^\dag$ instead of $\zeta_{n l \omega}$.

Given the rule adopted in calculating the square root of $\mathrm{i}\kappa$, the real component of eigenvalues is always non-positive and is closest to zero 
in the most symmetric case, $n=0$ and $l=0$. In the approximation we use, the absolute value of the decrement

\begin{equation}
-\mathrm{Re}\lambda_{\omega n l}=m\sqrt{\frac{\pi\nu}{2c}\omega\ln\Lambda}\left(1+|n|+2l\right), \label{decrement}
\end{equation}
and therefore, for a fixed $n, l$ mode it is proportional to the square root of the frequency. In particular, the decrement is zero at zero frequency --
the fact that simply reflects the conservation of energy, which is only redistributed between different light beams.

This dependence allows one to understand the physical meaning of this damping. Indeed, it comes from two approximations: 
first, we have neglected the potential time delays, 
which makes $\kappa\propto\omega^{-1}$; second, the incidence angles are small, and therefore geometric time delays are proportional to squares of these 
angles, resulting in second-order differential 
operator in $\btau$. Therefore, we are dealing with a partial loss of coherency in a given $\omega, n, l$ 
mode due to random geometric time delays uncorrelated between lensing planes. Relative importance of such delays is different for different $n, l$ modes, 
as described by~(\ref{decrement}).

\section{An illustration: the case of an isotropic source}

Let us now consider a test case where emission from a source located at $D=-D_S$ enters the deflectors-filled
region of space $D>0$. It is convenient to rewrite the initial intensity moment as 
\begin{equation}
M_0\left({\bf r}, \balpha, t\right)=f\left({\bf r}-\balpha D_s, \balpha, t-\frac{\balpha^2}{2c}D_s\right) , \label{iiim}
\end{equation}
because for an isotropic source, the function $f$ is independent of its second argument, and in the paraxial approximation used, 
isotropy can be assumed to describe most of the astrophysical sources. In this case, $f$ represents just a spatial structure
of the source and its variability $f({\bf x}, s)$, with ${\bf x}, s$ being the spatial coordinate and time at the
source position $-D_s$. Making Fourier transform of~(\ref{iiim}) yields\footnote{Regularization might be done by adding a small
negative constant to $\mathrm{i}\omega D_s/2c$, then letting it tend to zero}:
\begin{equation}
M_0\left({\bf k}, \btau, \omega\right)=f\left({\bf k}, \omega\right)\frac{\mathrm{i}c}{\omega D_s}\exp\left[\frac{-\mathrm{i}c}{2\omega D_s}\left(\btau + {\bf k}D_s\right)^2\right], \label{iiimko}
\end{equation}
where $f({\bf k}, \omega)$ is the source Fourier transform.


Projecting this initial conditions onto the basis vectors and summing the resulting series at $\btau=0$, one obtains for the flux:
\begin{equation}
F_{{\bf k}\omega}(D)=\frac{\mathrm{i}c f\left({\bf k}, \omega\right)}{\omega D_s}\frac{w(\omega, D)\exp[-\frac{\mathrm{i}c}{2\omega}{\bf k}^2\left(D_s+D\right)]}{\cos{\frac{\omega D}{c}\sqrt{\mathrm{i}\kappa}}+\frac{c}{\omega D_s\sqrt{\mathrm{i}\kappa}}\sin{\frac{\omega D}{c}\sqrt{\mathrm{i}\kappa}}} . \label{iiimfluxsolution}
\end{equation}
This is to be compared to the propagation of flux in the absence of deflectors, $\kappa=0$:
\begin{equation}
F_{{\bf k}\omega}(D)/F_{{\bf k}\omega}(D)|_{\kappa=0}=w(\omega, D)K(\omega, D), \label{effect}
\end{equation}
which defines the `filter function' $K(\omega, D)$. It can be expressed as a function of two parameters, $D/D_s$ and
$x\equiv\sqrt{\mathrm{i}\kappa}\omega D/c$:
\begin{equation}
K(\omega, D)=\left(\frac{\sin x}{x}+\frac{\cos x -\sin x / x}{1+D/D_s}\right)^{-1}\ \label{filter}
\end{equation}
and tends to an even simpler form
\begin{equation}
K_0(x)=\frac{x}{\sin{x}} \label{filter0}
\end{equation}
as $D\gg D_s$.

\begin{figure}
\hspace{0mm}
\vspace{0mm}
\includegraphics[width=80mm, angle=0]{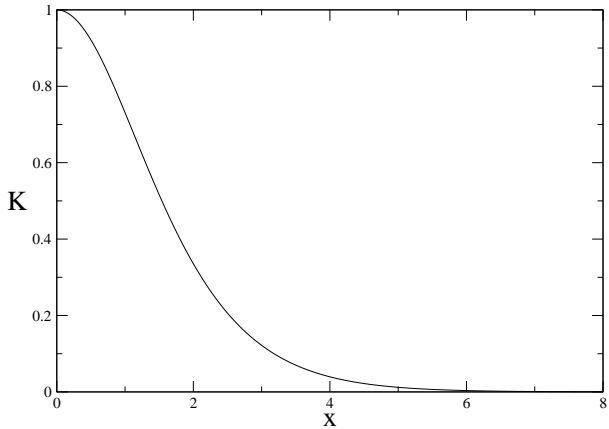}  
\caption{Minimum of the mode decay amplitude $\left|K_0\right|_{\mathrm{min}}$ as a function of $x_0$.}
\label{figK0min}
\hspace{0mm}
\vspace{0mm}
\end{figure}

Using~(\ref{kappadef}) we can separate the dependence of $x$ on the frequency
\begin{equation}
x=\sqrt{\pi\nu D^2 m\ln\Lambda}\sqrt{e^{\mathrm{i}\omega/\omega_0}-1} \label{xomega}
\end{equation}
with $\omega_0^{-1}= m\ln\Lambda/c$, so that the first factor $x_0=\sqrt{\pi\nu D^2 m\ln\Lambda}$ determines the amplitude 
of the mode decay due to lensing while the second sets the frequency scale: the modulus of $K_0$ reaches its minimum
\begin{equation}
\left|K_0\right|_{\mathrm{min}}(x_0)=\frac{2x_0}{\sqrt{\cosh{2\sqrt{2}x_0}-1}} \label{filteramplitude}
\end{equation}
at $\omega=\pi\omega_0$, and at low freqencies $\omega\ll\omega_0$ one has
\begin{equation}
\left|K_0\right|= 1 - \frac{15x_0^2+x_0^4}{180}\left(\frac{\omega}{\omega_0}\right)^2 + \mathcal{O}\left(\frac{\omega}{\omega_0}\right)^4. \label{filtersmallomega}
\end{equation}
The graph of $\left|K_0\right|_{\mathrm{min}}$ as a function of $x_0$ is plotted in figure~\ref{figK0min}.

\begin{figure}
\hspace{0mm}
\includegraphics[width=80mm, angle=0]{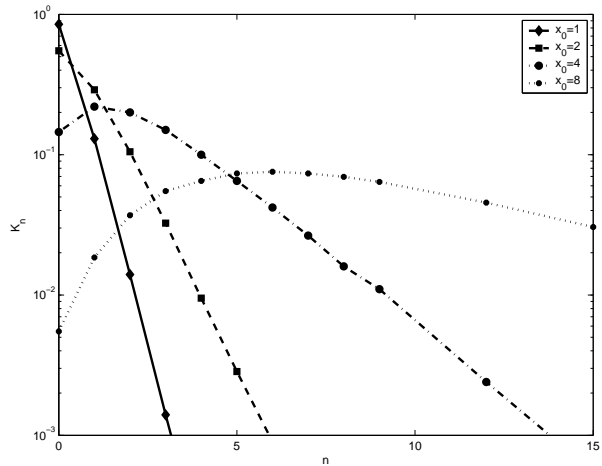}
\caption{The effect of applying the filter $K_0\left(x_0, \omega\right)$ to a single impulse at $t=0$. The dots show, for
a number of values of $x_0$, the magnitudes of the resulting peaks $K_n$; these are joined by the lines of constant $x_0$ for convenience purpose only as $K(x_0, t)=0$ at $t\not= nt_0$. 
The time scale unit is $t_0=2\pi\omega_0^{-1}=2\pi\ln\Lambda\, m/c$.}
\label{figK0}
\vspace{0mm}
\end{figure}

The effect of this filter is, perhaps, easiest to illustrate in the case of a single $\delta$-like in $t$ impulse. According
to the convolution theorem, this will be transformed into the inverse Fourier transform of $K_0(x_0, \omega)$. Since the filter function is strictly periodic in frequency, $K_0(x_0, t)$ will only be non-zero at $t= n t_0$, $t_0=2\pi\omega^{-1}_0=2\pi\ln\Lambda\,m/c$ ($n\in\mathcal{N}$); this
discreteness, however, is clearly an artefact of the approximation used for the deflection potential. Figure~\ref{figK0} presents the magnitudes
of these peaks $K_n(x_0)$ for a number of different values of $x_0$.

One should bear in mind that this broadening is calculated as the average over an ensemble. In relating this to
a single realization it can be argued that since microlensing is known to produce many microimages of the source -- each with
its own magnification and time delay -- the result in figure~\ref{figK0} should resemble an average shape of a microlensed 
$\delta$-like impulse. However, such an interpretation comes from outside the model and cannot be justified within it (see \citealt{williamswijers}, however, where
the results of a similar 2-D numerical calculation look surprisingly similar in terms of the time scale and the shape of the average lensed pulse in general).
A legitimate statement would be that the average light curve produced by a number of $\delta(t-t_s)$-like impulses {\it all occurring at the same moment} $t_s$
will be represented by $K_n$ (or a similar continuous function). It is hard to think of a physical reason for
such synchronization, though, and marginalizing over $t_s$ averages the effect to zero.

Similar consideration apply to damping of any harmonic modes which can be present in the initial signal. The average of
any mode amplitude over the ensemble -- which is the only quantity the approach of this paper allows to calculate consistently -- 
is always zero unless the initial phases of the oscillations in all individual realizations comprising the ensemble are somehow synchronized.
This is why we are inclined to conclude that this damping is even less likely to be observable in the first order than the broadening 
discussed above. Second order moments such as the power spectra, which do not suffer from the absence of
phase coherency are considered in the companion paper.

\section{Observational prospects}

To determine where the effect described in the previous section can be important -- regardless of the interpretation -- let us consider its amplitude 
parameter $x_0$. For observationally noticeable mode damping one need $x_0\ga 1$ as seen in the figure~\ref{figK0min}. There are
two conceivable ways to achieve this: by increasing $\nu$ or $D$. The former possibility corresponds to observing
a source through an extended overdensity along the line of sight; restricting ourselves to a single such
overdensity, we would interpret $D$ as the spatial extent of this structure rather than
the total path length towards the source. If the total number of the deflectors in this overdensity is fixed, and
it is largely spherically symmetric, one has $\nu\sim N D^{-3}$ and therefore
\begin{equation}
x_0\sim \sqrt{\frac{Nm}{D}\ln\Lambda}. \label{x0compact}
\end{equation}
As $Nm$ is (twice) the gravitational radius of the entire overdensity, it is unlikely that $x_0$ reaches a value
of order unity. For instance, it has been argued, that the $\mathrm{Sgr}\,\mathrm{A}^\star$ source in the centre of
the Galaxy could be a very dense cluster of stellar-mass ($10\,\mathrm{M}_\odot$) black holes instead of a single supermassive one, 
although this idea is almost certainly ruled out (\citealt{meliafalcke, meliacoker}). In this -- highly contrived --
case $Nm/D\sim 0.1$, $x_0$ becomes a quantity of order unity ($\ln\Lambda=\ln{D/m}\sim 10 - 100$) and
gravitational lensing could be an important effect when interpreting variability of objects seen behind the cluster
at frequencies close to $\omega_0\sim 10^3\,\mathrm{Hz}$. For more realistic conditions, however, $x_0$ would be
much lower and the effect can be ignored.

The other way to increase $x_0$ is to probe larger, cosmological distances along an on average uniform distribution 
of compact objects. One can rewrite $x_0$ in more relevant cosmological terms then:
\begin{equation}
x_0=\frac{DH}{c}\sqrt{\frac{3}{2}\Omega_d\ln\Lambda} , \label{x0cosm}
\end{equation}
where $\Omega_d$ is the deflectors' mass density in terms of the critical density and $H$ is the Hubble constant.
This expression cannot be precise since the analysis we presented in the previous sections was developed with stationary
three-dimensional geometry in mind, but it should give a fair estimate for moderate redshifts $z\la 1$. 

Therefore, for any appreciable cosmological population of compact objects $x_0$ reaches a value of order unity for
moderate $z$ and therefore such a population would manifest itself in suppressing high-frequency variability of distant
sources. The characteristic frequency, however, remains very high for reasonable values of the compact objects' masses:
as $\ln\Lambda\sim 100$ and $c/m=10^5(M/\mathrm{M}_\odot)^{-1}\,\mathrm{Hz}$, leading to $\omega_0\approx 10^3(M/\mathrm{M}_\odot)\,\mathrm{Hz}$.
At moderate redshifts, this rapid variability is likely beyond the reach of the present-day observational facilities. 

The only kind of sources which are bright enough to be probed at this frequency level seem to be the Gamma-Ray Bursts, 
but they are not expected to contain much harmonic signal in their intrinsic light curves at any frequencies. Being 
the results of multiple violent collisions between the shock waves in the expanding shells (e.g. \citealt{grbreview}), GRB 
light curves do not have the phase coherency necessary in the first-order effect considered in this paper. The question
of how their variability is affected by lensing can therefore be answered by the appropriate account of the second order
statistics -- e.g., auto-correlation functions or power spectra, which is the subject of the companion paper.

\section{Conclusions and discussion}

In the present paper, we have attempted to clarify the ideology used when studying how statistical properties of light from
distant objects are affected by microlensing. The important steps of the associated calculations which are often omitted and
the justification for various approximations which are normally assumed tacitly have been listed. We hope this will serve 
better understanding and attract further interest to the field which, over the past decades, has accumulated enough data to
make the statistical analysis of light curves routine.

In particular, we demonstrate how the conservation of the surface brightness leads to a particularly simple -- linear -- type 
of transformations when the statistical properties are defined in the phase space and one talks about intensity field rather
than fluxes. Connection of the phase space description with the observable flux, however, can only come through the average values of the
latter, and one has to increase the dimensionality of the problem in order to preserve linearity when studying higher moments
of the flux distribution. This is why we believe that the proposed approach will be more useful in analytic studies rather
than numerical ones. 

We distinguish between the two types of transformations corresponding to free propagation of light rays and
their deflection on lens planes because these have rather different properties and it is of advantage to represent them 
in different bases. The former, focusing, is not stochastic and in a sense, serves as a means of referencing the undeflected
light rays. Deflection, which, under the thin lens and small angle approximations, assumes a diagonal form in angular Fourier
domain, can have both the stochastic and deterministic parts and due to the linearity of the transformation these can be applied
separately.

For the stochastic part, we calculate the Fourier transform of the deflection angle and potential time delay probability 
density $p(\btau, \omega)$ thus extending the work by Katz et al. (1986). The potential time delays turned to be unimportant in observationally
interesting frequency range, however, and $p$ itself can be well represented by the first non-trivial term of its Taylor series expansion.

The method is also particularly suitable to study the case of many lens planes along the line of sight. To illustrate this, 
we take the multiple lens plane treatment to the extreme deriving an equation for the continuous propagation of intensity moments
under the assumption that the thin lens approximation still applies to the average quantities. The equation is solved for the 
first order moments affected by randomly distributed compact objects and the results are presented with an emphasis on 
the propagation of the flux temporal variability. We have devoted a fair fraction of the paper to the solution of the equation
because it will form the basis to study the second order moments.

The conclusion we reach is that when astrophysically reasonable parameters 
of microlens populations are assumed, the effect on the average amplitudes of harmonic oscillations starts to be important 
at frequencies which cannot be effectively probed in observations at the present time, and the variability expected from
astrophysical sources does not extend to those high frequencies. This is not the case for the second moments as we show in 
the second paper of this series for the autocorrelation function.

The scope for further development of the method includes, first of all, generalization to non-static and, perhaps, non-flat
geometry which would make possible its application to cosmologically significant distances beyond $z\sim 1$; this is also
likely to eliminate some awkward inconsistencies in the calculation of the deflection probability Fourier transform such as
the use of rather arbitrary cut-off radius. Another important direction is to obtain the solutions for more accurate approximations to the exact transformation
potential $V$ as well as derive an expression for this function for deflectors other than the point-like one. Account of
correlations between deflectors position is also of great interest since massive objects in the real Universe are all 
clustered; deflector proper motions also need to be considered and we do so in the companion paper for a Gaussian 
velocity distribution.

The continuous approximation presented in this paper is not, however, a truly three-dimensional approach as the thin lens
approximation is still in use and it is not clear how 3-D clustering of deflectors can be incorporated in this treatment.
What the true three-dimensional statistical lensing will look like, therefore, remains to be seen.

\section*{Acknowledgements}
AVT is supported by IPRS and IPA from the University of Sydney and thanks Mark Walker, Dmitry Klochkov, Daniil Smirnov,
Don Melrose and Rodrigo Gil-Merino for very valuable discussions and numerous important suggestions. We thank the referee
for their constructive criticism which helped us to clarify the presentation significantly.

\appendix

\section[]{Calculation of the deflection probability Fourier transform}
\label{potentialappendix}

To calculate the deflection probability Fourier transform we expand~(\ref{Vfromp1}) in Taylor series according to~(\ref{pseries}) 
and then calculate the integral~(\ref{expectationphi}) in ${\bf r}'$, using~(\ref{phidisc}):
\begin{eqnarray}
\lefteqn{p(\btau, \omega)=\frac{1}{\pi R^2}\times} \label{pdiscseries} \\
& & \int\limits_0^{2\pi}\mathrm{d}\theta\int\limits_\epsilon^R\mathrm{d}\rho\, \rho\left\{1+\sum\limits_{n=1}^\infty\frac{1}{n!}\left(\mathrm{i}m\left(\omega s_0(\rho)+\btau\cdot\bbeta_0(\brho)\right)\right)^n\right\}\nonumber
\end{eqnarray}
With a change $x\equiv\rho/R$, $\epsilon_x\equiv\epsilon/R$, $x_0\equiv\rho_0/R$, $a\equiv-m\omega/c$ and ${\bf b}\equiv\btau m/R={\bf e}_\theta\tau m/R$ one has
\begin{eqnarray}
\lefteqn{p(\btau, \omega)-1=\int\limits_0^{2\pi}\frac{\mathrm{d}\theta}{\pi}\int\limits_{\epsilon_x}^1\mathrm{d}x\, x\sum\limits_{n=1}^\infty\frac{\mathrm{i}^n}{n!}
\left(a\ln\frac{x}{x_0}-b\frac{\cos\theta}{x}\right)^n} \label{pidiscscaled} \\
& & \hspace{9mm} = -\mathrm{i}\frac{2m\omega}{c}\int\limits_{\epsilon_x}^1\mathrm{d}x\, x\ln\frac{x}{x_0} \nonumber\\
& & \hspace{9mm} + \int\limits_0^{2\pi}\frac{\mathrm{d}\theta}{\pi}\int\limits_{\epsilon_x}^1\mathrm{d}x\, x\sum\limits_{n=2}^\infty\frac{\mathrm{i}^n}{n!}
\left(a\ln\frac{x}{x_0}-b\frac{\cos\theta}{x}\right)^n \nonumber
\end{eqnarray}
We choose $x_0$ such as to eliminate the second line. This quantity represents an overall
potential time delay due to a smooth matter distribution associated with $\phi(\brho)$ and given a certain freedom to
define the time axis, we set the factor at $m\omega/c$ to zero. This leads to
\begin{equation}
x_0\approx\exp[-\frac{1}{2}-\epsilon_x^2\ln\epsilon_x]\approx e^{-1/2}\sim 1 \label{x0def}
\end{equation}
or $\rho_0\approx R/\sqrt{e}\sim R$.

One then further expands the $n$-th power of the expression in parentheses of~(\ref{pidiscscaled}) using Newton's binomial formula
\begin{equation}
\left(a\ln\frac{x}{x_0}-b\frac{\cos\theta}{x}\right)^n=\sum\limits_{k=0}^nC^k_n\left(-b\frac{\cos\theta}{x}\right)^k\left(a\ln\frac{x}{x_0}\right)^{n-k}\label{binom}
\end{equation}
and $p-1$ becomes
\begin{eqnarray}
\lefteqn{p(\btau, \omega)-1=} \label{psum}\\
& & \sum\limits_{n=2}^\infty\frac{\mathrm{i}^n}{n!}\sum\limits_{k=0}^n\int\limits_0^{2\pi}\mathrm{d}\theta\,\frac{\cos^k\theta}{\pi}\int\mathrm{d}x\,x\left(-\frac{b}{x}\right)^k\left(a\ln\frac{x}{x_0}\right)^{n-k}\nonumber
\end{eqnarray}
Only $k$-even terms survive integration with respect to the angular coordinate producing
\begin{equation}
\Xi_k\equiv\frac{1}{\pi}\int\limits_0^{2\pi}\mathrm{d}\theta\,\cos^k\theta=2\frac{(k-1)!!}{k!!}, \hspace{2mm} k=2j,\label{Xikdef}
\end{equation}
while for all $k$-odd terms $\Xi_{2j-1}=0$.

In radial integration, four rather different types of integrals arise:
\begin{eqnarray}
\lefteqn{k=n,\hspace{15mm} I^n_n\equiv\int\limits_{\epsilon_x}^1\frac{\mathrm{d}x}{x^{n-1}},} \label{Inndef}\\
\lefteqn{4\le k < n,\hspace{9mm} I^k_n\equiv\int\limits_{\epsilon_x}^1\frac{\mathrm{d}x}{x^{k-1}}\,\ln^{n-k}\frac{x}{x_0},} \label{Ikndef}\\
\lefteqn{k=2,\hspace{15mm} I^2_n\equiv\int\limits_{\epsilon_x}^1\frac{\mathrm{d}x}{x}\,\ln^{n-2}\frac{x}{x_0},} \label{I2ndef}\\
\lefteqn{k=0,\hspace{15mm} I^0_n\equiv\int\limits_{\epsilon_x}^1\mathrm{d}x\,\ln^n\frac{x}{x_0},} \label{I0ndef}
\end{eqnarray}
which will be estimated separately. Two are elementary:
\begin{equation}
I^n_n=\frac{\epsilon_x^{2-n}-1}{n-2}\approx\frac{\epsilon_x^{2-n}}{n-2} ,\hspace{16mm}n\ge 4 \label{Inn}
\end{equation}
and
\begin{eqnarray}
\lefteqn{I^2_n=\frac{1}{n-1}\left(\ln^{n-1}\frac{1}{x_0}-\ln^{n-1}\frac{\epsilon_x}{x_0}\right)}\nonumber \\
& & \approx(-1)^n\frac{\ln^{n-1}x_0/\epsilon_x}{n-1}, \hspace{19mm} n\ge 2. \label{I2n}
\end{eqnarray}
For $4\le k<n$ the integral is strongly dominated by the lower integration limit as $\epsilon_x\ll 1$, therefore
\begin{equation}
I^k_n\approx (-1)^{n-k}\frac{\ln^{n-k}x_0/\epsilon_x}{(k-2)\epsilon_x^{k-2}} \, \hspace{15mm} 4\le k< n, \label{Ikn}
\end{equation}
and one can see that at this level of approximation~(\ref{Inn}) is just a special case of~(\ref{Ikn}), so the last inequality
can be changed to an non-strict one.

For $k=0$ there are two distinct regimes: for large $n\ga\bar{n}\equiv 2\ln x_0/\epsilon_x$
\begin{equation}
I^0_n\approx\frac{(-1)^n\epsilon_x^2}{n+1}\ln^{n+1}\frac{x_0}{\epsilon_x}\, , \label{I0nlargen}
\end{equation}
while for smaller $n$ $I^0_n$ should be computed by direct integration or using a different approximation
\begin{equation}
I^0_n\approx\frac{(-1)^nx_0^2}{2^{n+1}}\left(n!-2^n\frac{\epsilon_x^2}{x^2_0}\ln^n\frac{x_0}{\epsilon_x}\right)\label{I0nsmalln}\, .
\end{equation}
We will write $I^0_n=\bar{I}^0_n+\Delta I^0_n$ with $\bar{I}^0_n$ defined by the right-hand side of~(\ref{I0nlargen}) and
$\Delta I^0_n$ becoming unimportant for $n\ga\bar{n}$.

Rearranging summation order in~(\ref{psum}) one obtains:
\begin{eqnarray}
\lefteqn{p(\btau, \omega)-1=\sum\limits_{n=2}^\infty\sum\limits_{j=0}^{[n/2]}\frac{\mathrm{i}^n}{n!}C^{2j}_nI^{2j}_n\Xi_{2j}a^{n-2j}b^{2j}} \label{psumrearr}\\
& &\hspace{9mm}=2\sum\limits_{n=2}^\infty\frac{(\mathrm{i}a)^n}{n!}I^0_n+\frac{1}{2}\left(\frac{b}{a}\right)^2\sum\limits_{n=2}^\infty\frac{(ia)^n}{(n-2)!}I^2_n \nonumber\\
& &\hspace{12mm} +\sum\limits_{j=2}^\infty\frac{b^{2j}\Xi_{2j}}{(2j)!}\sum\limits_{n=2j}^\infty\frac{\mathrm{i}^na^{n-2j}}{(n-2j)!}I^{2j}_n \nonumber \, .
\end{eqnarray}
Plugging expressions for $I^k_n$ above into these sums produces:
\begin{eqnarray}
\lefteqn{p(\btau, \omega)-1 = \epsilon_x^2f_\omega(a, \epsilon_x)+\Delta f_\omega(a,\epsilon_x)}\label{pfinal} \\
& & \hspace{10mm} -\frac{b^2}{2}e^{-\mathrm{i}a\ln x_0/\epsilon_x}\left[\frac{e^{\mathrm{i}a\ln x_0/\epsilon_x}-1}{ia}+f_\tau\left(-\frac{b^2}{4\epsilon_x^2}\right)\right] ,  \nonumber
\end{eqnarray}
where
\begin{equation}
f_\tau(t)\equiv\frac{1}{2}\sum\limits_{m=1}^\infty\frac{t^m}{m[(m+1)!]^2}=\int\mathrm{d}t\frac{I_0(2\sqrt{t})-t-1}{2t^2}\label{ftaudef} \, ,
\end{equation}
\begin{eqnarray}
\lefteqn{f_\omega(a, \epsilon^{-1}_x)\equiv \frac{1+\mathrm{i}a\ln x_0/\epsilon_x-a^2/2\ln^2x_0/\epsilon_x-e^{-\mathrm{i}a\ln x_0/\epsilon_x} }{\mathrm{i}a} }\nonumber\\
& &\hspace{6mm} \approx -\frac{a^2}{6}\ln^3\frac{x_0}{\epsilon_x} \label{fomegadef}
\end{eqnarray}
and
\begin{equation}
\Delta f_\omega(a, \epsilon^{-1}_x)\equiv\sum\limits_{n=2}^\infty\frac{(\mathrm{i}a)^n}{n!}\Delta I^0_n \label{dfomegadef}
\end{equation}
should be well represented by its first $\sim\bar{n}(\epsilon_x)$ terms. To the first order of approximation it is just
$-a^2x_0^2/8$.

Returning to original variables, one obtains ($\Lambda=R/m$):
\begin{eqnarray}
\lefteqn{ p(\tau, \omega)\approx 1+\frac{m^2}{R^2}f_\omega\left(\frac{m\omega}{c}, \Lambda\right) +\Delta f_\omega\left(\frac{m\omega}{c}, \Lambda\right) } \label{ptauomega} \\
& & \hspace{0mm} -\frac{m^2\tau^2}{2R^2}e^{\mathrm{i}\frac{m\omega}{c}\ln\Lambda}\left[f_\tau\left(-\frac{\tau^2}{4}\right)-\frac{\mathrm{i}c}{m\omega}\left(1-e^{-\mathrm{i}\frac{m\omega}{c}\ln\Lambda}\right)\right], \nonumber
\end{eqnarray}
from which the expression~(\ref{Vdisc}) follows immediately.

Finally, we note two approximations. First, using~(\ref{fomegadef},\ref{dfomegadef}) one obtains for the first line of~(\ref{Vdisc}) at low frequencies:
\begin{eqnarray}
\lefteqn{V(\tau, \omega)=-\pi\nu\frac{m^2\omega^2}{8c^2}\left(R^2+\frac{4}{3}m^2\ln^3\Lambda\right)+V'(\tau,\omega) .} \label{Vsmallomega}
\end{eqnarray}


\begin{figure}
\hspace{0cm}
\includegraphics[width=80mm, angle=0]{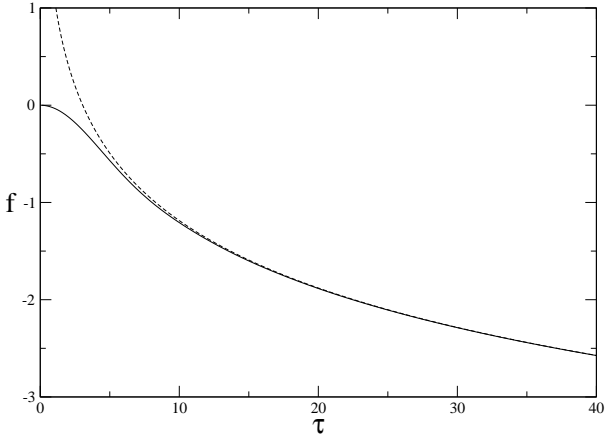}
\caption{Function $f_\tau\left(-\frac{\tau^2}4\right)$, defined by~(\ref{ftaudef}) (solid line) and approximation~(\ref{ftauasymptotic}) (dashed line).}
\label{ftaufig}
\end{figure}

Second, for large negative arguments $f_\tau$ has logarithmic behaviour $f_\tau(-x)\approx 1-\tilde{\gamma}-\frac{1}{2}\ln x$,
where $\tilde{\gamma}\approx 0.577216$ is Euler's gamma constant, therefore
\begin{equation}
f_\tau\left(-\frac{\tau^2}{4}\right)\approx 1-\tilde{\gamma}+\ln 2 -\ln\tau \label{ftauasymptotic} \, ,
\end{equation}
as shown in Fig.~\ref{ftaufig}. Thus, it is a slowly varying function, and the potential, to good accuracy, is quadratic in $\btau$.




\section[]{Basis of solutions for propagation equation}
\label{solutionappendix}
In this appendix we build a complete system of solutions of the equation~(\ref{xipproblem}) by further separation of $D$ and $\btau$ variables:
\begin{equation}
\xi'(D; \btau)=\sum\limits_\lambda e^{\lambda D}\xi'_\lambda(\btau) \label{eigensum}
\end{equation}
replacing the original problem with
\begin{equation}
\lambda\xi'_\lambda={\bf k}\cdot\partial_\btau\xi'_\lambda-\frac{\mathrm{i}\omega}{2c}\partial^2_\btau\xi'_\lambda + V'\xi'_\lambda \label{eigenxip}
\end{equation}
for the set of all possible $\lambda$. For $\omega \not = 0$, which is of primary interest for this study, it is more
convenient to divide both sides by $-\mathrm{i}\omega/2c$ denoting $\lambda=-\mathrm{i}\omega\mu/2c$ and get rid of ${\bf k}$-proportional term 
by further transforming
\begin{equation}
\xi'_\lambda(\btau)=\zeta'_\mu(\btau)\exp\left[-\mathrm{i}\frac{c}{\omega}{\bf k}\cdot\btau\right] \label{zetapdef}
\end{equation}
to obtain the following equation for $\zeta'_\mu(\btau)$:
\begin{equation}
\mu\zeta'_\mu(\btau)=\partial^2_\btau\zeta'_\mu(\btau)+\left(\mathrm{i}\frac{2c}{\omega}V'(\btau, \omega)+\frac{c^2}{\omega^2}{\bf k}^2\right)\zeta'_\mu(\btau) \label{zetapeq}
\end{equation}
Now, introducing cylindrical coordinates $\btau=\tau{\bf e}_\chi$, and separating variables again, while observing periodicity 
\begin{equation}
\zeta'_{n\mu}(\tau{\bf e}_\chi)=\zeta_{n\mu}(\tau)e^{-\mathrm{i}n\chi}, \hspace{10mm}n\in{\mathcal Z}, \label{zetadef}
\end{equation}
we arrive at an ordinary differential equation
\begin{equation}
\mu\zeta_{n\mu}(\tau)=\frac{1}{\tau}\frac{\mathrm{d}}{\mathrm{d}\tau}\left(\tau\frac{\mathrm{d\zeta_{n\mu}(\tau)}}{\mathrm{d}\tau}\right)+v_n(\tau)\zeta_{n\mu}(\tau) \label{odezeta}
\end{equation}
with
\begin{equation}
v_n(\tau)\equiv\mathrm{i}\frac{2c}{\omega}V'(\tau, \omega)+\frac{c^2}{\omega^2}{\bf k}^2-\frac{n^2}{\tau^2} \label{vndef}
\end{equation}
It is clear that the term depending on $k$ can be absorbed into $\mu$, so for each ${\bf k}$ mode the resulting equation has
the same form with $\mu_{{\bf k}}=\mu-\frac{c^2}{\omega^2}{\bf k}^2$. We will see soon that for an oscillator-like potential additional constraints on $\zeta_{n\mu}$ lead to a discrete set of $\mu$ like in the corresponding QM case.

Neglecting $f_\tau$ term in $V'(\btau)$ as explained in Section 3 and Appendix~\ref{potentialappendix}, one rewrites the potential as
\begin{equation}
v_n(\tau)=-\left(\frac{n^2}{\tau^2}+\mathrm{i}\kappa\left(\omega, \tau\right)\tau^2\right)\label{oscillator}
\end{equation}
with
\begin{equation}
\mathrm{i}\kappa(\omega)=\pi\nu m^3 \frac{\exp[\mathrm{i}\omega/\omega_0]-1}{(\omega/\omega_0)^2} \approx\mathrm{i}\pi\nu\frac{m^2c}\omega\ln\Lambda  . \label{kappadef}
\end{equation} 
Note that $\mathrm{i}\kappa(\omega, \tau)=\left(\mathrm{i}\kappa(-\omega, \tau)\right)^*$; this guarantees that
the solution to the problem with real initial conditions remains real for $D>0$.


We thus arrive at the differential equation
\begin{equation}
\mu_{\bf k}\zeta_{n\mu}(\tau)=\frac{1}{\tau}\frac{\mathrm{d}}{\mathrm{d}\tau}\left(\tau\frac{\mathrm{d}\zeta_{n\mu}}{\mathrm{d}\tau}\right) - \left(\frac{n^2}{\tau^2}+\mathrm{i}\kappa(\omega)\tau^2\right)\zeta_{n\mu}(\tau); \label{oscillatorode}
\end{equation}
we will drop ${\bf k}$ and $\omega$ indices for now. The differential part of the equation 
above is Hermitean with respect to the standard weight in problems with cylindrical symmetry ($\mathrm{d}\tau\,\tau$). 
However, since the oscillator constant is not real, the total operator is not Hermitean leading to different modes decaying 
at different rates. Therefore, we also need to solve the equation conjugate to~(\ref{oscillatorode}):
\begin{equation}
\sigma\zeta^\dag_{n\sigma}(\tau)=\frac{1}{\tau}\frac{\mathrm{d}}{\mathrm{d}\tau}\left(\tau\frac{\mathrm{d}\zeta^\dag_{n\sigma}}{\mathrm{d}\tau}\right) - \left(\frac{n^2}{\tau^2}-\mathrm{i}\kappa^*\tau^2\right)\zeta^\dag_{n\sigma}(\tau) \label{conoscillatorode}
\end{equation}
solutions to which will form a conjugate set to the set of solutions of~(\ref{oscillatorode}):
\begin{equation}
(\zeta^\dag_\sigma, \zeta_\mu)=0, \hspace{1cm} \sigma^*\not = \mu. \label{orthogonality}
\end{equation}

Our aim is to represent the initial conditions as a linear combination of the solutions to~(\ref{oscillatorode}) in the most 
economic way; the coefficients of this linear combination are found using solutions to~(\ref{conoscillatorode}). In fact, 
this `non-Hermitecity' is easily eliminated by a simple change of the independent variable and the solutions to~(\ref{oscillatorode}) and~(\ref{conoscillatorode})
can be obtained from those of a single equation. Indeed, if $u_r(z)$ equates
\begin{equation}
ru_r(z)=\frac{1}{z}\frac{\mathrm{d}}{\mathrm{d}z}\left(z\frac{\mathrm{d}u_r(z)}{\mathrm{d}z}\right)-\left(\frac{n^2}{z^2}+z^2\right)u_r(z) \label{hermode}
\end{equation}
then it is easy to check that $\zeta_r(\tau)=u_r(c\tau)$ and $\zeta^\dag_r(\tau)=u_r(c^*\tau)$ are the solutions of~(\ref{oscillatorode}, \ref{conoscillatorode}) with $\mu=rc^2$ and $\sigma=r^*(c^*)^2=\mu^*$, provided that
\begin{equation}
c^4=\mathrm{i}\kappa . \label{cdef}
\end{equation}

Here one rather technical but important observation concerning the choice of $c$ is due. The problem~(\ref{hermode}) does have a compact set of solutions $u_r(z)$ which can be used to represent any square-integrable on the real positive ray function $g(t)$, and all $r$ in this set are real. Therefore, given a function $f(\tau)$ we can always represent it as
a linear combination of $\zeta_\mu$ if $f(\tau/c)$ is square-integrable on $\tau\in(0, +\infty)$. In all cases of interest, the 
required convergence only takes place at $\mathrm{Re}\, z >0$ half of the complex plane, therefore among the four roots of~(\ref{cdef}) we will choose ones with real positive part (which of the two is relatively unimportant as long as it is used consistently, we will use the one closest to the real axis because such choice is generally less restrictive on $f(\tau)$). To distinguish the appropriate $c$ we will use the square root notation $c^2=\sqrt{\mathrm{i}\kappa}$, square root being understood in the above, arithmetic, sense. Let us also note that for real $r$~(\ref{orthogonality}) changes into
\begin{equation}
(\zeta^\dag_r, \zeta_{r'})=0, \hspace{1cm} r\not = r'. \label{realorthogonality}
\end{equation}

To solve the equation~(\ref{hermode}) we first change the dependent variable
\begin{equation}
u(z)=v_{\pm}(z) z^{\pm n}\exp\left[-\frac{z^2}2\right] \label{change}
\end{equation}
and then a change of the independent variable $v(z)=w(z^2)$ brings it to the confluent hypergeometric form:
\begin{equation}
t w_{\pm}''(t)+(1\pm n-t)w_{\pm}'(t)-\left(\frac{1\pm n}2+\frac{r}4\right)w_{\pm}(t)=0, \label{confode}
\end{equation}
where prime denotes the derivative with respect to the function $w$ argument $t$.

The two independent solutions of the last equation are given by the confluent hypergeometric 
functions of the first and second kinds:
\begin{eqnarray}
\lefteqn{w_{\pm}(t)=C_M M\left(\frac{1\pm n}2+\frac{r}4, 1 \pm n, t\right)} \label{genconf}\\
& & \hspace{1cm} + C_U U\left(\frac{1\pm n}2+\frac{r}4, 1 \pm n, t\right), \nonumber
\end{eqnarray}
$C_M, C_U$ are complex numbers.

The solutions to~(\ref{hermode}) are therefore
\begin{equation}
u^{\pm}_M(z)=z^{\pm n}\exp\left[-\frac{z^2}2\right]M\left(\frac{1\pm n}2+\frac{r}4, 1 \pm n, z^2\right)
\end{equation}
and
\begin{equation}
u^{\pm}_U(z)=z^{\pm n}\exp\left[-\frac{z^2}2\right]U\left(\frac{1\pm n}2+\frac{r}4, 1 \pm n, z^2\right)
\end{equation}
Of course, only two of these four are linearly independent as can be readily established \citep{AS13}. The most universal choice
of the two is $u_M(z)=z^{|n|}\exp\left[-\frac{z^2}2\right]M\left(\frac{1+|n|}2+\frac{r}4, 1+|n|, z^2\right)$ 
and $u_U(z)=z^{|n|}\exp\left[-\frac{z^2}2\right]U\left(\frac{1+|n|}2+\frac{r}4, 1+|n|, z^2\right)$. Again, only one of these
two, $u_M$ is regular at $\tau=0$ while $u_U$ has a pole of order $|n|$ here. Since we are not interested in singular
solutions, the solution of interest to us is
\begin{equation}
u_r(z)=z^{|n|}\exp\left[-\frac{z^2}2\right]M\left(\frac{1+|n|}2+\frac{r}4, 1+|n|, z^2\right) \label{soluhermode}
\end{equation}
Now we recall that $\mathrm{Re}\, \sqrt{\mathrm{i}\kappa(\omega)}=\mathrm{Re}\, \sqrt{-\mathrm{i}\kappa^*(\omega)}>0$ and from
the asymptotics of the confluent hypergeometric function \citep{AS13} it follows that $\zeta_r$ and 
$\zeta^\dag_r$ grow exponentially with $\tau\rightarrow\infty$ unless the first argument of $M$ is
a real, non-positive, integer number $-l$, $l=0, 1, 2, ...$ in which case $M(-l, 1+|n|, t)$ becomes a polynomial of
degree $l$ in $t$ known as the associated Laguerre polynomial of order $|n|$ and degree $l$ $L^{|n|}_l(t)$.

These polynomials form a basis in $\mathcal{L}_2(\mathcal R^+)$ and therefore can be used
to represent any initial conditions $M^0_{{\bf k} \omega n}(\tau)$ of interest. This solves the problem: the complete eigensystem of~(\ref{oscillatorode}) within regular at zero $\mathcal{L}_2(\mathcal{R}^+)$ functions is
\begin{eqnarray}
\lefteqn{\zeta_l(\tau)=c_l\tau^{|n|}\exp\left[-\frac{\sqrt{\mathrm{i}\kappa}}2\tau^2\right]L^{|n|}_l\left(\sqrt{\mathrm{i}\kappa}\tau^2\right)} \\\nonumber
\lefteqn{\mu_l=-2\sqrt{\mathrm{i}\kappa}\left(1+|n|+2l\right), \hspace{1cm} l=0, 1, 2,...\, .} \label{eigensystemzeta}
\end{eqnarray}
while the complex conjugates of the above function are the complete eigensystem of~(\ref{conoscillatorode}): $\zeta^\dag_l=\zeta^*_l$, 
$\sigma_l=\mu^*_l$.

It is convenient to normalize them to unity $(\zeta^\dag_l, \zeta_{l'})=\delta^l_{l'}$, which implies
\begin{equation}
c_l=\sqrt{\frac{2l!}{(|n|+l)!}\left(\mathrm{i}\kappa\right)^{(|n|+1)/2}}; \label{normzeta}
\end{equation}
here we used that $(L^{|n|}_l(t))^2 t^n\exp[-t]$ is regular between $\arg t= \arg\sqrt{\mathrm{i}\kappa}$ 
and $\arg t=0$, the integral over the arc connecting these rays approaches zero as its radius goes to infinity as well
as orthogonality relations from \cite{GR}.

The solution to the original problem~(\ref{smek}, \ref{initialconditions}) 
is therefore given by
\begin{equation}
M(D)=\sum\limits_{{\bf k}nl\omega}w(\omega, D) \mathrm{e}^{\lambda_{{\bf k}nl\omega}D} \left(M^\dag_{{\bf k} n l \omega}, M_0\right) M_{{\bf k}nl\omega} \label{smeksolution}
\end{equation}
with the basis functions, their conjugates and corresponding eigenvalues being
\begin{eqnarray}
\lefteqn{M_{{\bf k}nl\omega}\left(\tau, \chi \right)=} \label{fullmodes} \\
& & \hspace{0.5cm}\sqrt{\frac{l!\sqrt{\mathrm{i}\kappa}}{\pi\left(|n|+l\right)!}} \left(\left(\mathrm{i}\kappa\right)^{1/4}\tau\right)^{|n|}L^{|n|}_l\left(\sqrt{\mathrm{i}\kappa}\tau^2\right) \nonumber \\
& & \hspace{0.5cm} \times\exp\left[-\mathrm{i}\left(n\chi+\frac{c}\omega k\tau\cos(\chi-\phi)\right)-\sqrt{\mathrm{i}\kappa}\frac{\tau^2}{2}\right], \nonumber
\end{eqnarray}
\begin{eqnarray}
\lefteqn{M^\dag_{{\bf k}nl\omega}\left(\tau, \chi \right)=} \label{confullmodes} \\
& & \hspace{0cm}\sqrt{\frac{l!\sqrt{-\mathrm{i}\kappa^*}}{\pi\left(|n|+l\right)!}} \left(\left(-\mathrm{i}\kappa^*\right)^{1/4}\tau\right)^{|n|}L^{|n|}_l\left(\sqrt{-\mathrm{i}\kappa^*}\tau^2\right) \nonumber \\
& & \hspace{0cm} \times\exp\left[-\mathrm{i}\left(n\chi+\frac{c}\omega k\tau\cos(\chi-\phi)\right)-\sqrt{-\mathrm{i}\kappa^*}\frac{\tau^2}{2}\right] \nonumber
\end{eqnarray}
and 
\begin{equation}
\lambda_{{\bf k}nl\omega} =-\frac{\mathrm{i}c}{2\omega}{\bf k}^2+\frac{\mathrm{i}\omega}{c}\sqrt{\mathrm{i}\kappa(\omega)}\left(1+|n|+2l\right). \label{fullvalues}
\end{equation}
Here ${\bf k}=k{\bf e}_\phi$, $\btau=\tau{\bf e}_\chi$ and $w(\omega, D)$ is defined by~(\ref{wdef}).


\section[]{Approximation for the spread function}
\label{spreadappendix}
As discussed in the section 4, function $w(\omega, D)$ defined by~(\ref{wdef}) represents the Fourier transform of the
`spread function' $w(t-t', D)$ with which one has to convolve the solution of problem~(\ref{xipproblem}). This spread
function is universal in the sense that it is the same for all angular modes and thus physically describes variance
in the potential time delay within the ensemble of deflector field realizations.

The inverse Fourier transform of $w(\omega)$ simplifies in the case when parameters of the problem (such as $\nu$, $R$, $m$) are constant,
so that $w(\omega, D)=\exp[\pi\nu D (R^2\Delta f_\omega+m^2 f_\omega)]$. In addition, most of the variability
we are concerned with is concentrated at $\omega$ range where $\omega/\omega_0\ll 1$ and $w(\omega)$ is approximately Gaussian, according to~(\ref{fomegadef}, \ref{dfomegadef}):
\begin{equation}
w(\omega)\approx \exp\left[-\pi\nu D\frac{m^2\omega^2}{8c^2}\left(R^2+\frac{4}{3}m^2\ln^3\Lambda\right)\right] \label{Gaussspread}
\end{equation}
The inverse transform of this function is again a Gaussian 
\begin{equation}
w(t-t')=\left(2\pi\sigma^2_t\right)^{-1/2}\exp\left[-\frac{(t-t')^2}{2\sigma^2_t}\right]\label{Spreadint}
\end{equation}
with dispersion
\begin{equation}
\sigma^2_t(D)=\frac{m^2}{4c^2}\nu DR^2\left(1+\frac{4\ln^3\Lambda}{3\Lambda^2}\right) \, , \label{sigmatdef}
\end{equation}
and given that the time it takes light to traverse gravitational radius of the deflector of mass $m$ is just $t_m=m/2c$, and
the total number of deflectors is $N(D)=\pi\nu R^2D$, the effective width of the spread function $w$ is
\begin{equation}
\sigma_t(D)=t_m\sqrt{N}\sqrt{1+\frac{4\ln^3\Lambda}{3\Lambda^2}}\approx t_m\sqrt{N} \label{spread}
\end{equation}
The result might seem somewhat surprising in a sense that this dispersion depends on the total number of deflectors regardless
of their location with respect to the observer in the final plane. However, the analysis we develop in this work does not 
involve an observer up to the point when some interpretation of the result is needed. Indeed, the variance above is 
a quantity defined on the plane labelled $D$ (or, more precisely, the ensemble average of such planes) and therefore one 
needs access to the entire plane in order to measure it.

When the observer can sample only a fraction of the plane, the dispersion will be less; moreover, for uncorrelated 
deflectors it is generally proportional to the square root of the number of deflectors -- but only those which influenced 
the rays which have been observed. Our conjecture is therefore that in calculating~(\ref{spread}) one only needs the number 
of deflectors within the region sampled by the observer over the run of the data set used. 
Here we touch 
the question of how average quantities measured on a distribution sample relate to those of the distribution itself, which 
is a complex problem on its own studied in statistics. A natural way to quantify the uncertainty of such measurements is 
provided by considering their second moments. This is the subject of the companion paper.

\label{lastpage}
\end{document}